\documentclass[12pt,preprint]{aastex}

\def\ga{\mathrel{\raise0.35ex\hbox{$\scriptstyle >$}\kern-0.6em
\lower0.40ex\hbox{{$\scriptstyle \sim$}}}}
\def\la{\mathrel{\raise0.35ex\hbox{$\scriptstyle <$}\kern-0.6em
\lower0.40ex\hbox{{$\scriptstyle \sim$}}}}

\shorttitle{Molecular gas in HE\,0450-2958}
\shortauthors{Papadopoulos, Feain, Wagg, \& Wilner}
\begin{document}

\title{A new twist to an old story: HE\,0450-2958, and the
ULIRG$\rightarrow $(optically bright QSO) transition hypothesis}

\author{Padeli \ P.\ Papadopoulos}
\affil{Argelander-Institut f\"ur Astronomie,  Auf dem H\"ugel 71,  D-53121 Bonn, Germany}
\email{padeli@astro.uni-bonn.de}

\author{Ilana J. Feain}
\affil{CSIRO Australia Telescope National Facility, P.O. Box 76, Epping, NSW 1710, Australia}
\email{Ilana.Feain@CSIRO.AU}

\author{Jeff Wagg}
\affil{NRAO, PO Box 0, Socorro, New Mexico, 87801,USA}
\email{jwagg@nrao.edu}

\and

\author{David J. Wilner}
\affil{Harvard-Smithsonian Center for Astrophysics, Cambridge, MA, 02138, USA}
\email{dwilner@cfa.harvard.edu}


\begin{abstract}

We report on interferometric imaging  of the CO J=1--0 and J=3--2 line
emission from the  controversial QSO/galaxy pair HE\,0450--2958.  {\it
The  detected CO  J=1--0 line  emission is  found associated  with the
disturbed  companion galaxy not  the luminous  QSO,} and  implies $\rm
M_{gal}(H_2)\sim (1-2)\times 10^{10}\,  M_{\odot}$, which is $\ga 30\%
$ of  the dynamical  mass in its  CO-luminous region.  Fueled  by this
large gas  reservoir this galaxy is  the site of  an intense starburst
with $\rm  SFR\sim 370\, M_{\odot  }\, yr^{-1}$, placing it  firmly on
the  upper  gas-rich/star-forming   end  of  Ultra  Luminous  Infrared
Galaxies  (ULIRGs,  $\rm L_{IR}>10^{12}\,  L_{\odot  }$).  This  makes
HE\,0450--2958 the  first case of  extreme starburst and  powerful QSO
activity, intimately  linked (triggered  by a strong  interaction) but
not coincident.  The lack of CO emission towards the QSO itself renews
the controversy regarding its host  galaxy by making a gas-rich spiral
(the  typical  host  of  Narrow  Line  Seyfert~1  AGNs)  less  likely.
Finally,  given  that  HE\,0450--2958  and similar  IR-warm  QSOs  are
considered   typical   ULIRG$\rightarrow   $(optically   bright   QSO)
transition candidates, our results raise the possibility that some may
simply  be  {\it  gas-rich/gas-poor (e.g.   spiral/elliptical)  galaxy
interactions} which ``activate'' an optically bright unobscured QSO in
the gas-poor  galaxy, and a starburst  in the gas-rich  one.  We argue
that such  interactions may  have gone largely  unnoticed even  in the
local  Universe   because  the  combination  of   tools  necessary  to
disentagle the progenitors (high  resolution and S/N optical {\it and}
CO imaging) became available only~recently.

\end{abstract}


\section{Introduction}

 HE\,0450-2958  is an optically  bright ($\rm  M_v=-25.8$) IR-selected
 QSO at a redshift of z=0.286 (Low et al. 1988), and one of the few in
 the  local Universe  whose position  in  the (60$\mu  $m, 25$\mu  $m)
 versus (100$\mu $m, 60$\mu $m) far-IR color-color diagram lies in the
 area between Ultra Luminous  Infrared Galaxies (ULIRGs) and optically
 bright QSOs defined  by $\rm -0.8 < \alpha (100, 60)  < 0.4$ and $\rm
 -2.0 <  \alpha (60, 25)  < - 0.8$  (Canalizo \& Stockton  2001).  Its
 complex  environment, marked  by tidal  interactions,  was identified
 early  with ground-based  (Hutchings,  \& Neff  1988)  and {\it  HST}
 imaging  (Boyce et  al.  1996).   These efforts  revealed  a strongly
 interacting system consisting of the QSO and a disturbed galaxy $\sim
 1.5''$ away, setting this QSO/galaxy  pair on par with typical ULIRGs
 in terms  of morphology, and  IR luminosity ($\rm  L_{IR}\sim 5\times
 10^{12}\, L_{\odot}$).

The  scarcity  of such  IR  {\it  and}  optically luminous  QSOs  made
 HE\,0450-2958   an  early   favorite  candidate   for   undergoing  a
 ULIRG$\rightarrow $(optically  bright QSO) transition  (Hutchings, \&
 Neff 1988; Canalizo \& Stockton 2001) in a scenario first outlined by
 Sanders  et   al.   (1988).   The  latter  was   proposed  after  FIR
 color-color diagrams have  identified a population of dust-enshrouded
 QSOs via  their AGN-heated  dust component (de  Grijp, Miley,  \& Lub
 1987), and  it was motivated  by the similar  bolometric luminosities
 and  space densities  of optically  powerful QSOs  and ULIRGs  in the
 local  Universe.   It involves  two  gas-rich  galaxies whose  strong
 dynamical  interaction induces  a  starburst and  fuels  an AGN  {\it
 within} heavily dust-obscured  environments.  Then, as star formation
 uses  and  disperses  their  large molecular  gas  reservoirs,  their
 initially cool  Spectral Energy  Distribution (SED) of  dust emission
 changes  towards one  dominated  by warm  AGN-heated  dust with  much
 warmer  IR   colors.   Eventually  an   unobscured  optically  bright
 QSO~emerges out of the original  ULIRG.  The large dust and molecular
 gas reservoirs found in {\it optically} selected QSOs where they fuel
 intense starbursts  (e.g.  Alloin  et al.  1992;  Haas et  al.  2000;
 Evans  et al.  2001)  certainly support  an AGN-starburst  link, with
 dynamical  interactions/mergers  as   the  likely  trigger  for  both
 activities,  suggested also  by  optical studies  (e.g.  Canalizo  \&
 Stockton   2001).   In  this   context  the   pivotal  role   of  the
 ULIRG$\rightarrow $QSO  transition objects can then  be understood as
 the ``markers'' of a special  evolutionary stage during which the QSO
 emerges  from a  still ongoing  dusty starburst,  and lasting  only a
 fraction of already short  molecular gas consumption timescales ($\rm
 \la 10^8\,yrs$).

HE\,0450-2958  acquired further singular  significance when  Magain et
al.   (2005), after  carefully subtracting  the AGN  emission  from an
ACS/HST  image, failed  to  find  the QSO  host  galaxy expected  from
well-established  and  tight   correlations  of  (host  galaxy)-quasar
properties (e.g.   McLure \& Dunlop  2002; Floyd et al.   2004).  This
potentially  pivotal result  has been  upheld by  independent analysis
(Kim  et  al.  2007),  and  follow-up  VLT spectroscopic  observations
weakened the  hypothesis for a significantly  dust-enshrouded QSO host
galaxy  (Letawe  et  al.    2007).   The  possibility  of  an  ejected
(black-hole)+(fuel)  system as  responsible for  a ``naked''  QSO, its
serious consequences and observable  signatures have been discussed in
several papers.  Summarizing their  results here, such a configuration
can be produced either by a Newtonian three-body ``kick'' exerted on a
lighter  black hole  by a  black hole  binary residing  in  a gas-poor
elliptical (Haehnelt, Davies, \& Rees  2005; Hoffman \& Loeb 2006), or
by the recoil of a coalesced  binary of {\it spinning} black holes due
to an  asymmetric gravitational wave emission (Haehnelt  et al.  2005;
Loeb  2007).  The  stage for  both  scenaria demands  a strong  galaxy
interaction,  and both have  important consequences  for gravitational
wave  detections by  future  detectors  such as  {\it  LISA}, and  the
hierarchical build-up  of supermassive  black holes in  galaxy centers
(Haehnelt et al.  2005 and references~therein).

In the  present work we report  on the observations of  CO J=1--0, the
prime molecular line  used to trace metal-rich H$_2$  gas in galaxies,
using the  Australian Telescope Compact  Array (ATCA), and  the J=3--2
line  with the Smithsonian  Submillimeter Array  (SMA).  The  paper is
organized  as  follows:  a)  we  present the  observations  and  their
analysis (section 2),  b) estimate the molecular gas  and dust mass of
the system and discuss the properties of the companion galaxy (Section
3),  c)  present the  implications  regarding  the  AGN and  its  host
(section   4),  and   d)  discuss   possible  ramifications   for  the
ULIRG$\rightarrow  $QSO evolutionary  scheme (section  5).  Throughout
this     work      we     adopt     a      cosmology     with     $\rm
H_{\circ}=71\,km\,s^{-1}\,Mpc^{-1}$,  $\rm  \Omega  _M=0.27$ and  $\rm
\Omega  _{\Lambda}=0.73$,   for  which  the   luminosity  distance  of
HE\,0450-2958  at  z=0.286  is   $\rm  D_{L}=1460.4\,  Mpc$  and  $\rm
1''\rightarrow 4.28\,kpc$.

\section{Observations, data reduction, and imaging}

\subsection{The ATCA observations}

We  used the  ATCA to  image  the CO  J=1--0 line  emission ($\rm  \nu
 _{rest}=115.2712\, GHz$) towards  the QSO/galaxy system HE\,0450-2958
 during  four periods  in  April,  August 2006  and  May, August  2007
 utilizing the  hybrid H\,214 and H\,168  configurations which combine
 antennas on  both the  east-west track and  north track  with maximum
 baselines   of  168\,m  and   214\,m  respectively.    These  special
 configurations  provide good  brightness sensitivity  while achieving
 good u-v  coverage within  short ($\sim $6\,hrs)  observing sessions.
 This  is  important  for  mm  observations at the  ATCA  site  where
 atmospheric conditions usually limit  useful observing (i.e. with low
 system  temperatures  and good  phase  stability)  to time  intervals
 significantly  shorter than  typical full  synthesis  sessions ($\sim
 $12\,hrs).   For the  mean  redshift $\rm  z  = 0.2864$  of the  pair
 (Canalizo \& Stockton 2001), this CO transition is redshifted to $\rm
 \nu (z)= 89.608\, GHz$, well  within the tuning range of ATCA's 3\,mm
 receivers (available in  5 of the 6 antennas).   The correlator setup
 consisted of two IF modules  of $\rm 64 \times 2\,MHz$ channels each,
 and frequency resolution of $\rm \Delta \nu _{res}\sim 2.2 \Delta \nu
 _{ch}=4.4\, MHz  $.  The effective velocity coverage  was $\rm \Delta
 V_{(IF1+IF2)}\sim   (-300\rightarrow  +270)\,  km\,   s^{-1}$,  after
 flagging bad edge channels  and channel overlap (=7 channels) between
 the two IF modules.  Pointing was checked hourly, and typical offsets
 were $\sim  5''$ (rms), while the  field of view of  ATCA antennas at
 this frequency is  32$''$ (HPBW).  The array phase  center was placed
 at the  AGN's position, marked by  its radio core at  8.6~GHz (the C1
 component in Feain et al.  2007).

  Dedicated wideband  mm continuum observations  were conducted during
 periods in May, August 2006, and  August 2007 using the two IFs tuned
 in series around a center frequency $\rm \sim 94.53\, GHz$ (well away
 from the CO line), and the same array configurations (two tracks with
 H\,214, three tracks with  H\,168).  For these observations the phase
 center was  placed $3''$ away  from the QSO  as to avoid  any (highly
 unlikely)  small correlator DC  offsets masking  as a  weak continuum
 source at the phase  center.  Typical system temperatures during line
 and  continuum  observations   were  $\rm  T_{sys}\sim  (200-350)\,K$
 (including  atmospheric  absorption),   estimated  with  hourly  vane
 calibration performed per~antenna.

Amplitude/phase  calibration was  obtained with  0454-234 observations
interleaved  with  HE\,0450-2958  in 1min/(3-5)mins  calibrator/source
intervals, and passband calibration was achieved by observing 1921-293
or 2223-052.  The absolute flux density scale was set by bootstrapping
that of 0454-234  from Uranus and/or Mars observations  in each track.
These yielded $\rm  S_{3\,mm}(0454-234)=(1.9\pm 0.28)\, Jy$, where the
$\sim 15\%$ error  represents the flux scale uncertainty  of the data,
obtained as the dispersion of $\rm S_{3\,mm}(0454-234)$ over the 2006,
2007 observing periods  (and thus is an upper  limit since it contains
also any real source~variability).

\subsection{The SMA observations}

A  search  for  the  redshifted  CO J=3--2  line  emission  ($\rm  \nu
 _{rest}=345.796\,   GHz$)   was   conducted   using   the   8-element
 SMA\footnote{The Submillimeter  Array is a joint  project between the
 Smithsonian  Astrophysical   Observatory  and  the   Academia  Sinica
 Institute  of  Astronomy  and  Astrophysics  and  is  funded  by  the
 Smithsonian Institution and  the Academia Sinica.}  interferometer on
 9th of December 2006 with  a median atmospheric opacity at 225~GHz of
 $\tau   _{225}\sim   0.08$.   This   transition   is  redshifted   to
 269.102~GHz,  covered by  the  lower sideband  (LSB)  of the  345~GHz
 receivers placed at 268.892 GHz,  while the upper sideband was offset
 by 10 GHz from this  center frequency.  The bandwidth covered by each
 is 2  GHz while the spectral  resolution was 3.25~MHz for  a total of
 768 channels.   This was smoothed so  that a single CO  J=3-2 map was
 created,  averaged over  a  velocity range  of  570 $\rm  km\,s^{-1}$
 centered at the redshift of the CO J=1-0 line emission.

The  SMA extended  configuration was  used, yielding  a beam  of $\sim
1.20'' \times  0.94''$, PA=16.9$^\circ$ at the  observed frequency and
the  u-v coverage  attained.   The mean  system  temperature was  $\rm
T_{sys}=490\,K$ (DSB), and the  total on-source integration time $\sim
5.07$ hrs.   Calibration of  the antenna gain  variations was  done by
observing the calibrators 0522-364  and 0455-462, at regular 10~minute
intervals.  Saturn  and 3C454.3 were  used to calibrate  the bandpass,
while  Uranus was  observed  for flux  calibration,  with an  expected
uncertainty  of $\sim$20\%.   The MIR  package was  then used  for the
calibration of  the visibility~data.   A search for  275~GHz continuum
yielded  an  upper  limit   of  $\rm  S_{275\,GHz}\leq  8\,  mJy/beam$
($3\,\sigma$),   over  a   $\rm  \sim   3.936\,GHz$   bandwidth  (both
sidebands), while the noise in the line map was $\rm \sigma _{rms}\sim
6.6\,mJy/beam$.

\subsection{Data reduction and imaging}

We used  the {\it  in-situ} ATCA phase  monitor operating on  a 230\,m
east-west baseline  (Middelberg, Sault, \& Kesteven  2006) to identify
the periods when atmospheric phase noise at 90\,GHz over this baseline
(comparable  to the longest  present in  our datasets)  exceeded $\sim
30^{\circ}$ (rms)  and rejected the  data taken during  those periods.
The remaining  visibilities were then  edited for any  remaining large
temporal  amplitude/phase jumps.   The final  $\rm  V(u,v)$ visibility
dataset has  a residual phase rms of  $\sigma _{\phi}\sim 20^{\circ}$,
in accordance with  a mean coherence factor $\rm  \rho _{coh} =\langle
V(u,v)\rangle/S =e^{-\sigma  ^2 _{\phi}(rad)/2} \sim  0.95$, estimated
from imaging  the visibilities  of the calibrator  with an  input flux
density  S (and  antenna  amplitude/phase solutions  smoothed to  $\rm
T_{avg}  \sim   2\times  T_{cycle}(QSO\leftrightarrow  calibrator)\sim
12\,mins$ intervals).  The u-v range of the final dataset is $\rm \sim
(12\rightarrow 75)\, k{\lambda}$ for both line and continuum.

Imaging was done  using MIRIAD's task INVERT with  natural weight used
  for maximum point-source sensitivity in our line and continuum maps.
  CO J=1--0 emission was detected independently in both the H\,214 and
  H\,168 array  datasets, with peak flux densities  differing by $\sim
  15\%$, consistent  with the calibration uncertainties.   An image of
  the   CO   J=1--0   emission   was   produced   by   multi-frequency
  synthesis\footnote{Fourier    transform   of   the    $\rm   V(u,v)$
  visibilities {\it per channel}, then averaging the resulting channel
  maps}  over  the  whole  band, and  combining  both  configurations.
  Deconvolution  using the  Clark algorithm  was then  performed.  The
  resulting CO map and mm continuum image of the same region are shown
  in Figure 1, and the  spectrum corresponding to the peak CO emission
  is shown in Figure 2.

\subsubsection{Source characteristics}

The velocity-averaged CO J=1--0  image yields the highest S/N possible
from our data ($\rm S_{peak}/\sigma_{rms}\sim $15, Figure 1), and thus
allows optimum determination of the  position of the peak emission and
the  source size.   A gaussian  source model  yields an  excellent fit
(i.e. after subtraction of the  model, the noise in the residual image
over the source area is $\rm \sigma _{rms} =0.45\,mJy/beam$, as in the
rest   of  the   map  area)   where  the   peak  CO   brightness  $\rm
S_{peak}=(6.7\pm  0.45)\,  mJy/beam$ is  located  at $(\Delta  \alpha,
\Delta \delta)  = (1.15''\pm  0.12'', -1.05''\pm 0.10'')$  relative to
the AGN's radio core at the phase center.  The small positional errors
are those expected  from the high S/N ($\rm  \delta \theta _{rms} \sim
1/2\, \langle  \Theta _{beam} \rangle  (S/N)^{-1}\sim 0.09''$ relative
to the  phase center).  The  angular proximity of  the amplitude/phase
calibrator  to HE\,0450--2958 ($\rm  |\Delta \vec  k|\sim 6.5^{\circ}$
away) allows also for  excellent {\it absolute} astrometry by limiting
$\rm  \delta \theta  _{bas} =  (\delta  \vec B\cdot  \Delta \vec  k)/B
\approx (\delta \phi  _{bas}/2\pi )\langle \Theta _{beam}\rangle$ (the
uncertainty  due  to  the  phase  error $\rm  \delta  \phi  _{bas}\sim
(2\pi/\lambda) (\delta \vec B\cdot  \Delta \vec k)$) to $\sim 0.086''$
(for typical calibration errors of the baseline length of $\rm |\delta
\vec  B|\sim 1\,mm$).  {\it  Thus we  are confident  that the  peak CO
emission is not located at  the QSO position but $\sim 1.56''\pm0.17''
(\sim 6.7\,kpc)$ away.}

The total  flux density  of $\rm S  ^{(tot)} _{\nu }(CO)=  (10\pm 2)\,
 mJy$ emerges for  a source for which the  gaussian fit gives: $\theta
 _{maj}=3.95''  \pm 0.30'' $  and $\theta  _{min}=2.80'' \pm  0.22'' $
 (PA=$57^{\circ}\pm  9^{\circ} $).   Deconvolving  the restoring  beam
 ($3.21''\times 2.14''$, PA=$72^{\circ}$)  yields an intrinsic size of
 $2.5''\times  1.5''$  (PA=$35^{\circ}$),  larger  than  any  expected
 ``seeing''   disk.   The   latter,   deduced  from   images  of   the
 amplitude/phase calibrator, is  $\sim 0.30''-0.45''$ (consistent with
 the  expected   $\rm  \Delta  \theta   _{seeing}\sim  [\sigma  _{\phi
 }(rad)/2\pi]  \sqrt{8\,  ln2}\,  \langle \Theta  _{beam}\rangle  \sim
 0.34'' $,  for a  residual $\rm \sigma  _{\phi }\sim  20^{\circ}$ and
 $\rm  \langle \Theta  _{beam}\rangle  =\sqrt{\Theta _1\Theta  _2}\sim
 2.62''$).

No continuum emission is detected at 94.5\,GHz in either the starburst
 or the  QSO position ($\rm  S_{94\,GHz}<1.4\,mJy$ (3$\sigma$), Figure
 1).  All  relevant data, and the  characteristics of the  CO and dust
 emission in HE\,0450--2958 are summarized in Table 1.

\section{The molecular gas in HE\,0450--2958}

The CO J=1--0  emission overlaid with the 8.4\,GHz  radio continuum is
shown in  Figure~2 where  a good correspondence  with the  region C\,2
associated with the companion galaxy  (Feain et al.  2007) is evident.
Regriding the observed CO emission and its deconvolved model, and then
overlaying  them  onto the  HST/ACS  optical  frame of  HE\,0450--2958
further underlines  its association  with the companion  galaxy rather
than the QSO (Figure 4).

  The  corresponding  H$_2$  mass   is  given  by  $\rm  M(H_2)=X_{CO}
L_{CO(1-0)}$, where  $\rm L_{CO(1-0)}=  \int _{\Delta V}  \int_{A_s} T
_{b}\,dA\,dV  $   is  the  velocity/area-integrated   line  brightness
temperature at the  source reference frame, and $\rm  X_{CO}$ (in $\rm
M_{\odot  }  (K\,km\,s^{-1}\,   pc^2)^{-1}$  units)  is  the  CO-H$_2$
luminosity-mass conversion  factor.  Using standard  derivations (e.g.
Solomon et al.  1997) we have

\begin{equation}
\rm L_{CO(1-0)} =  3.25\times 10^7\,(1+z)^{-1} 
\left(\frac{\nu _{co,rest}}{GHz}\right)^{-2} \left(\frac{D_L}{Mpc}\right)^2
 \left(\frac{\int S_{CO(1-0)} dV}{Jy\,km\,s^{-1}}\right)\, K\, km\,s^{-1}\,pc^2.
\end{equation}

\noindent
For  the   observed  $\rm   \int  S_{CO(1-0)}\,dV$  (Table   1),  $\rm
D_L=1460.4\,Mpc$  and  $\rm \nu  _{co,rest}=115.271\,  GHz$ (the  line
rest-frame  frequency),  the  latter  gives  $\rm  L_{CO(1-0)}=(2.3\pm
0.46)\times 10^{10}\, K\,  km\, s^{-1}\, pc^2$.  This is  close to the
maximum CO luminosity observed for starbursts, beyond which it remains
constant  across redshift  (Frayer  et al.   1999;  Evans, Surace,  \&
Mazzarella 2000),  a possible indication of  a self-regulating process
occuring in extreme (maximal?)  starbursts.  Using $\rm X_{CO}\sim 1\,
M_{\odot}  (K\,km\,s^{-1}\,pc^2)^{-1}$   (deduced  for  local  ULIRGs,
Downes  \& Solomon  1998), we  obtain $\rm  M_{gal}(H_2)\sim 2.3\times
10^{10}\, M_{\odot}$,  {\it which places the companion  galaxy in this
galaxy/QSO pair  firmly in the  ULIRG category,} at its  most gas-rich
upper  end,  and  similar   to  starbursts  found  at  high  redshifts
(e.g. Greve et al.~2005).

A lower limit  in the H$_2$ gas mass can be  obtained by assuming that
the $ ^{12}$CO J=1--0 emission is optically thin. Using the derivation
in  Bryant \&  Scoville (1996)  and assuming  LTE, the  new conversion
factor becomes

\begin{equation}
\rm \frac{M(H_2)}{L_{CO(1-0)}} \sim 0.08 \left(\frac{[CO/H_2]}{10^{-4}}\right)^{-1}
\left[\frac{g_1}{Z}e^{-T_{\circ}/T_k} \left(\frac{J(T_k)-J(T_{bg})}{J(T_k)}\right)\right]^{-1} 
\frac{M_{\odot }}{K\,km\,s^{-1}\,pc^2},
\end{equation}

\noindent
where     $\rm    T_{\circ}     =     E_1/k_B\sim    5.5\,K$,     $\rm
J(T)=T_{\circ}\left(e^{T_{\circ}/T}-1\right)^{-1}$,                $\rm
T_{bg}=(1+z)T_{cmb}\sim  3.5\,K$ (CMB  temperature  at z=0.286),  $\rm
g_1=3$ (degeneracy factor of  n=1 level), $\rm Z\sim 2(T_k/T_{\circ})$
(partition  function),  and  $\rm  [CO/H_2]\sim  10^{-4}$  (for  solar
metallicity).   For  typical  star  forming  gas  where  $\rm  T_k\sim
40-60\,K$,   the  last   equation  yields   $\rm   \langle  X^{(thin)}
_{CO}\rangle        _{T_k}        \sim       0.55\,        M_{\odot}\,
(K\,km\,s^{-1}\,pc^2)^{-1}$,  and  thus  $\rm M_{gal}(H_2)  _{min}\sim
1.25\times 10^{10}\,M_{\odot }$.

\subsection{Constraints on the molecular gas excitation}

 The CO  J=3--2 observations  yield a limit  of $\rm  S_{CO(3-2)}\la 3
{N_b}^{1/2}   \delta  S_{CO(3-2)}\Delta   V  $,   where   $\rm  \delta
S_{CO(3-2)}\sim 6.6\,  mJy/beam $ is the  noise of the  CO J=3--2 map,
$\rm N_{b}\sim 3.3$ is the number of SMA beams corresponding to the CO
J=1--0 emitting  region and $\rm \Delta  V\sim 570\,km\,s^{-1}$.  This
gives   $\rm  S_{CO(3-2)}\la   20\,  Jy\,   km\,  s^{-1}$,   and  $\rm
L_{CO(3-2)}\la  9\times  10^{9}\,   K\,  km\,  s^{-1}\,  pc^2$,  which
corresponds to  a (3--2)/(1--0)  brightness temperature ratio  of $\rm
R_{32}\la 0.39$.  This  is rather low for molecular  gas in starbursts
where  $\rm \langle  R_{32}\rangle \sim  0.65$ (e.g.   Devereux  et a.
1994), though  within the range found  in such galaxies  (e.g.  Yao et
al.  2003).   We examined an extensive  grid of models  with our Large
Velocity  Gradient (LVG)  code  (based on  work  by Richardson  1985),
restricted by the upper limit  on $\rm R_{32}$, and the condition $\rm
T_{\rm  kin}\geq   T_{dust}\ga  50\,K$  (since  the   gas  cools  less
efficiently than  the dust, and  turbulent/photoelectric heating heats
the  gas more).   Most  conditions reproducing  this  ratio have  $\rm
\langle  n(H_2)\rangle   \sim  (10^2-10^3)\,cm^{-3}$  (volume-averaged
H$_2$ gas density),  rather low for star forming  H$_2$ gas.  This may
in turn signify  that not all of the  CO(1-0)-emitting gas is involved
in star formation (and thus  some of it is not CO(3-2)-bright).  Given
the minimum  $\rm R_{32}(min)\sim  0.22$ found in  ULIRGs (Yao  et al.
2003),  deeper SMA  observations of  CO J=3--2  at a  lower resolution
(better matching the CO J=1--0 source size) should detect this line in
HE\,0450--2958 and shed light on its star-forming molecular gas phase.

\subsection{The companion galaxy: a typical ULIRG}

The  mm/IR dust  continuum SED  of  HE\,0450--2958 from  IRAS and  our
mm/sub-mm data (Figure  5) is typical for warm  ULIRGs and IR-selected
QSOs, where the  presence of warm dust was the  prime reason they were
considered  as $\rm  ULIRG\rightarrow QSO$  transition objects  in the
evolutionary  scenario linking  these two  classes.  A  cool/warm dust
emission  SED fit  yields $\rm  T^{(cool)} _{dust}=45-55\,K$  and $\rm
T^{(warm)}  _{dust}\sim  175-194\,  K$  (for  emissivities  of  $\beta
=1-2$).   Adopting $\beta  =1.5 $  as  our working  value yields  $\rm
T^{(cool)} _{dust}=48\,K$, $\rm M^{(cool)} _{dust}\sim 10^8\, M_{\odot
}$, and $\rm L_{FIR}=2.1\times 10^{12}\, L_{\odot}$ (the luminosity of
the  cool dust  component).   A starburst  in  the gas-rich  companion
galaxy  is  a  natural  source  of this  large  far-IR  luminosity  of
HE\,0450-2958, with a star formation  rate of $\rm SFR \sim 1.76\times
10^{-10}(L_{IR}/L_{\odot})\,  M_{\odot}\,  yr^{-1}\sim 370\,  M_{\odot
}\,  yr^{-1}$,  and  a  star  formation  efficiency  of  $\rm  SFE\sim
L_{FIR}/M_{gal}(H_2)\sim  (90-165)\,  (L_{\odot}/M_{\odot})$,  typical
values  for ULIRGs (e.g.   Solomon et  al.  1997).   Moreover assuming
most of  the cool dust mass  in this system residing  in the companion
galaxy,  we obtain  $\rm  M_{gal}(H_2)/M^{(cool)}_{dust}\sim 125-230$.
These values are  well within the range found  in IR-luminous galaxies
(e.g.  Sanders Scoville, \& Soifer  1991, for the $\rm X_{CO}$ adopted
here), adding confidence to the estimated $\rm M_{gal}(H_2)$ as a good
measure of the bulk molecular gas mass present in HE\,0450--2958.

The  results above  are  certainly  in good  accord  with evidence  of
significant star formation and  reddening in the companion galaxy, and
no such  activity or dust obscuration  evident in a  putative QSO host
galaxy (Magain  et al.  2005;  Letawe, Magain, \& Courbin  2007). They
contradict the recent view of Kim et al.  2007 regarding the nature of
the companion  galaxy, but also  set HE\,0450-2958 apart  from typical
ULIRGs with double nuclei.  The  latter have molecular gas mass ratios
of $\sim 1:1-2:1$ (Sanders \& Ishida 2004; Evans, Surace, \& Mazzarela
2000),  and their  AGNs (when  present)  reside in  the most  gas-rich
member  of the  interacting/merger  pair (e.g.   Evans  et al.   1999,
2002).  Subtraction of  the CO emission model leaves  no residual such
emission   in  the   location  of   the  QSO   with   $\rm  S^{(peak)}
_{gal}/S_{QSO-host}\ga  5$ ($3\, \sigma$),  which corresponds  to $\rm
M_{gal}(H_2)/M_{QSO-host}(H_2) \ga  5:1$ (for a  common velocity range
and $\rm X_{CO}$~value).  A different (higher) $\rm X_{CO}$ factor for
the  molecular gas  in the  QSO host  could reduce  the aforementioned
asymmetry in  H$_2$ gas mass  distribution, while some CO  emission at
the QSO's location at velocities outside our velocity coverage of $\rm
\sim 570\,km\,s^{-1}$  could be missed.  However  the relatively small
velocity range traced by the  ULIRG, the QSO, and the highly excitated
ionized  gas  around  it  (Merritt  et  al.   2006)  make  the  latter
possibility rather small.  Sensitive wideband observations of e.g.  CO
J=3--2  with good  brightness sensitivity  would nevertheless  be very
valuable in the search for any QSO-related molecular gas.

\subsection{The dynamical  mass of the companion galaxy}

The  line  profile towards  the  peak CO  emission  shows  signs of  a
rotating disk (Figure~3), though the  lower S/N per channel across the
band  makes this far  from certain.   The CO  source model  shows good
overall agreement with  the galaxy's optical image (Figure  4), and at
$\rm  z=0.2864$  where  $\rm  2.5''\times  1.5''\rightarrow  10.7\,kpc
\times  6.4\, kpc$, its  dimensions are  comparable to  the H$_2$-rich
parts  of typical  spiral disks  (Regan et  al.  2001;  Helfer  et al.
2003), including the Milky Way, but with $\sim 10-20$ times more H$_2$
gas mass.

For  an  underlying  disk   geometry  the  source  angular  dimensions
correspond   to   an   inclination   angle  $\rm   cos   (i)=   \theta
_{minor}/\theta   _{major}=0.6\Rightarrow   i\sim  53^{\circ}$.    The
enclosed dynamical mass then would~be

\begin{equation}
\rm M_{dyn}\sim \frac{R}{G} \left(\frac{V^{(obs)} _{rot}}{sin(i)}\right)^2 \sim 2.32\times 10^5
\left(\frac{R}{kpc}\right)\left(\frac{V^{(obs)} _{rot}/sin(i)}{km\,s^{-1}}\right)^2\, M_{\odot}.
\end{equation}

\noindent
For   $\rm   R\sim    R_{major}/2\sim   5.35\,kpc$,   $\rm   V^{(obs)}
_{rot}=(100-150)\, km\, s^{-1}$ (see Figure 2), it is $\rm M_{dyn}\sim
(1.9-4.4)\times 10^{10}\, M_{\odot}$.  Thus the galaxy's molecular gas
mass  amounts to  $\ga 30\%$  of the  dynamical mass  enclosed  in its
CO-emitting region, making this ULIRG a very gas-rich system.

Inadequate resolution and/or low S/N makes it notoriously difficult to
discern a  gaseous disk from  other possible configurations,  and this
has  led  to ambiguities  regarding  the  presence  of large  gas-rich
spirals, especially in the distant  Universe (e.g. Genzel et al. 2003;
Tacconi et  al.  2006).  In our  case the resolution  is inadequate to
discern   velocity/position  patterns  within   the  source   and  the
underlying dynamical configuration  could certainly be different (e.g.
two orbiting compact gas  reservoirs).  Nevertheless for the intrinsic
size and  velocity width of  our source most  dynamical configurations
would  produce similar  or  even smaller  $\rm  M_{dyn}$ values  (e.g.
Bryant \& Scoville  1996; Genzel et al.  2003), and  thus the ULIRG in
HE\,0450-2958 would remain a very gas-rich~object in terms of its $\rm
M_{gal}(H_2)/M_{dyn}$ ratio.

\section{Implications for the QSO's host galaxy and its AGN}

   A  suggestion for  a Narrow  Line Seyfert~1  (hereafter  NLSy1) AGN
(which have  lower mass black holes),  made by Merrit  et al.  (2006),
could reconcile the lack of a massive spheroid host galaxy reported by
Magain et  al.  (2005).  However this  rests on the weak  premise of a
good analogy between luminous QSOs  and the much less luminous Seyfert
galaxies, unsupported by the fact that NLSy1 AGNs are typically hosted
by gas-rich spirals with luminous starbursts, while optically luminous
QSOs with  $\rm M_v<-24$ reside mostly in  gas-poor massive elliptical
hosts (Floyd  et al.  2004).  For example  in I Zw 1  (the closest QSO
and a prototypical NLSy1), a  spiral disk hosts a massive cirumnuclear
starburst at a radius of  $\rm \sim 1.9\, kpc$ (Schinnerer, Eckart, \&
Tacconi 1998;  Staguhn et al.  2004)  fueled by a  large molecular gas
reservoir.  {\it None of that is evident in the vicinity of the AGN in
the  HE\,0450-2958  system}  whose  bulk  of  the  molecular  gas  and
starburst  activity  are  found   in  the  companion  galaxy  instead.
Indicatively, at  z$\sim 0.286$ our CO 1--0  observations could detect
the molecular gas  mass of the $\sim 10$ times  less far-IR luminous I
Zw  1 at  a S/N$\sim  5$.  Finally,  well-established  {\it empirical}
AGN-(host  galaxy)   correlations  also  predict   a  hitherto  absent
prominent  elliptical (Magain  et al.~2005),  and thus  this important
issue remains~open.

\subsection{The AGN-related molecular gas: could it be ejected?}

The  two-component dust continuum  SED fit  (section 3.2)  yields $\rm
T^{(warm)}  _{dust}(AGN)=184\,K$,   $\rm  M^{(warm)}  _{dust}(AGN)\sim
5\times  10^4\, M_{\odot  }$ and  $\rm  L_{mid-IR}=2.6\times 10^{12}\,
L_{\odot}$ (the  luminosity of the corresponding  dust component), but
unlike the  typical scenario this  AGN-heated dust reservoir  now lies
{\it outside} the  ULIRG where the bulk of the  dust and molecular gas
mass  resides.  Using  the same  gas/dust  ratio as  in the  companion
galaxy  gives $\rm  M_{AGN}(H_2)\sim (0.6-1.15)\times  10^7\, M_{\odot
}$.  This  amounts to the mass of  a few GMCs, and  such molecular gas
quantities are  found within  $\rm \la 100\,pc$  of AGNs in  the local
Universe  (e.g. in  the Sy2  galaxy NGC  1068, Planesas,  Scoville, \&
Myers 1991; Schinnerer, Eckart, \& Tacconi~1999).

The  disturbed nature  of  the HE\,0450-2958  system  and the  extreme
starburst in  the gas-rich companion  galaxy $\sim 6.7$ kpc  away from
the QSO argues in favor of a strong dynamical interaction.  Could then
a  black hole,  along  with  the aforementioned  gas  mass, have  been
ejected out  of its  host galaxy  during such an  event?  In  order to
answer this  and check the  consistency of such proposals  against the
gas mass associated with this particular AGN, an estimate of its black
hole mass is first needed. A {\it lower} limit can be derived from the
fact  that the  dust within  an  AGN-bound gas  reservoir will  partly
obscure it (giving rise to the distinct warm IR colors), and thus

\begin{equation}
\rm L^{(AGN)} _{IR}= \frac{4\pi c G m_H\, f_c \epsilon _{Edd} M_{BH}}{\sigma _T}
= 3.3\times 10^{12}\, f_c\, \epsilon _{Edd} \left(\frac{M_{BH}}{10^8\,
 M_{\odot}}\right)\, L_{\odot}, 
\end{equation}

\noindent
where $\rm  f_c=L^{(AGN)} _{IR}/L_{AGN}\leq 1$ is the  fraction of the
AGN's intrinsic luminosity intercepted  by the dust and re-radiated at
IR wavelengths,  $\rm \epsilon _{Edd}=L_{AGN}/L_{Edd}$  ($\rm L_{Edd}$
is the Eddington  luminosity limit), $\rm \sigma _{T}$  is the Thomson
cross  section, and  $\rm M_{BH}$  the  black hole  mass.  Since  $\rm
L^{(AGN)}  _{IR}=2.6\times 10^{12}\,  L_{\odot }$,  the  last relation
yields,

\begin{equation}
\rm M_{BH}\sim 0.8 (f_c \epsilon _{Edd})^{-1}\, 10^8\,M_{\odot}.
\end{equation}

\noindent
For $\rm \epsilon _{Edd} \sim 0.8$ (typical for $\rm M_{V}(QSO)<-25 $,
Floyd  et  al.  2004),  and  $\rm  f_c=(1+R_{opt,IR})^{-1}\sim 0.71  $
(where  $\rm R_{opt,IR}=(L_{opt}/L_{IR})_{AGN}\sim  [\nu  _B f^{(AGN)}
_{\nu  }(B)]/[\nu_{60\mu  m}  f^{(AGN)}  _{\nu }(60\mu  m)]\sim  0.4$,
Canalizo  \&   Stockton  2001),   it  is  $\rm   M_{BH}\sim  1.4\times
10^8\,M_{\odot }$. This is consistent with the low values advocated by
Merritt et al.  (2006), but given  that ours is only a lower limit, it
does not  settle the  issue of the  black hole  mass (and thus  of the
expected host galaxy) raised by Magain et al.~(2005).

If such  a black hole  is the recoiling  remnant of a  coalesced black
hole binary then, following Loeb 2007,  it can carry a gas disk with a
mass of

\begin{equation}
\rm   M_{disk}\sim    3\times   10^5   \alpha^{-0.8}    \eta   ^{-0.6}
\left(\frac{M_{BH}}{10^7\,M_{\odot}}\right)^{2.2}
\left(\frac{V_{ej}}{10^3\, km\, s^{-1}}\right)^{-2.8}\, M_{\odot},
\end{equation}

\noindent
where $\alpha \sim  0.1$ is the disk viscosity  parameter, $\rm \eta =
(\epsilon  _{BH}/0.1)(L_{AGN}/L_{Edd})\sim 0.8$,  and $\rm  V_{ej}$ is
the ejection  velocity of  the recoiled merged  BH product.   For high
ejection velocities  of $\rm V_{ej}\ga  1000\,km\,s^{-1}$ (e.g Hoffman
\& Loeb 2006) Equation 6 yields the maximum $\rm M_{disk}$ that can be
carried out  by the recoiling  black hole of $\rm  M_{disk}\la 2\times
10^6\left[M_{BH}/(10^7\,M_{\odot})\right]^{2.2}\,  M_{\odot}$.   Given
that  $\rm   M_{BH}\ga  10^8\,M_{\odot  }$,   the  latter  comfortably
encompasses   the   molecular  gas   associated   with   the  AGN   in
HE\,0450-2958.

\subsubsection{How long could the AGN remain luminous?}

Apart  from partly  obscuring the  AGN in  HE\,0450-2958 via  its dust
content,  the molecular  gas mass  in  its vicinity  can in  principle
``fuel'' it for

\begin{equation}
\rm T_{fuel}=\frac{M_{AGN}(H_2)}{L _{AGN}/(c^2\,\epsilon _{BH})}\sim
1.4\times 10^8\, \epsilon _{BH}\, f_c \, \left(\frac{L^{(AGN)} _{IR}}{10^{12}\,L_{\odot}}\right)^{-1}
\left[\frac{M_{AGN}(H_2)}{10^7\,M_{\odot}}\right]\, yrs,
\end{equation}

\noindent
where $\rm \epsilon _{BH}$ is the black hole fuelling efficiency.  For
a  typical  $\rm\epsilon  _{BH}\sim  0.1  $  and  the  estimated  $\rm
M_{AGN}(H_2)$ the  latter yields $\rm  T_{fuel}\sim (2-4)\times 10^6\,
yrs$. The shortness of this timescale, and the difficulty of fueling a
black  hole  once  it  is  outside  its  host  galaxy,  may  point  to
fundamental  limitations  in  the  observability  of  such  events,  a
plausible reason for their hitherto scarcity.

\subsection{The origin of the radio continuum}

The  system's radio  continuum contains  low-luminosity  jets emerging
from  a  radio-quiet  core  located  at the  QSO  position  (Feain  et
al. 2007).  An important question arising from that study is how could
the  location  of  HE\,0450-2958  near the  FIR/radio  correlation  be
explained if the  starburst occurs in the companion  galaxy while most
of the radio  emission is of AGN origin.  Clues  for a possible answer
may be offered by the case of 1821+643, another radio-quiet but IR and
optically luminous  QSO ($\rm  M_V\sim -28$, $\rm  L_{FIR}\sim 9\times
10^{12}\, L_{\odot }$)  residing in a giant elliptical.   This was the
first radio-quiet QSO discovered  to have classic FR~I radio structure
with low  brightness diffuse jets  (Papadopoulos et al.  1995),  and a
high  brightness temperature  compact  core (Blundell  et al.   1996),
indicative of a black hole--based central engine.  Despite the obvious
AGN contribution in its radio  continuum it did not deviate decisively
from  the  far-IR/radio correlation  until  the  discovery  of a  much
larger, 300  kpc-sized, low-brightness FR I  radio structure (Blundell
\& Rawlings  2001).  A similar  scenario may apply  for HE\,0450-2958,
and the discovery  of such a jet morphology in  this system could also
offer independent clues regarding the black hole.

Interestingly,  FR  I-type  radio  structures  may  originate  from  a
precessing  jet  axis due  to  binary black  holes  that  have yet  to
coalesce  (Blundell \&  Rawlings  2001).  Such  binaries, residing  in
gas-poor  ellipticals (which  allows for  their long  survival against
gas-dynamical  friction),  are   a  fundamental  prerequisite  in  the
scenario of  ejected AGNs as the  outcome of a  three-body kick during
gas-rich/gas-poor galaxy interactions (Hoffman \& Loeb 2006).

\section{HE\,0450-2958: not an ULIRG$\rightarrow $QSO transition object, implications }

The  hereby  discovery  of  HE\,0450-2958,  considered  an  archetypal
 ULIRG$\rightarrow $QSO  transition object, as  a strongly interacting
 pair of a  gas/dust-poor galaxy (marked by a  residing unobscured and
 optically powerful  QSO) and a gas-rich extreme  starburst raises the
 prospect of {\it such interactions being misclassified in the context
 of the popular ULIRG$\rightarrow $QSO evolution scenario.}  More such
 cases  would then  signify  an important  variation  in the  standard
 evolutionary  picture  in which  starburst  and  QSO activity,  while
 triggered by a common  cause (a strong galaxy interaction/merger), do
 not  necessarily  involve two  gas-rich  progenitors with  comparable
 amounts   of  molecular   gas   (Sanders  \&   Ishida  2004).    Such
 gas-poor/gas-rich  galaxy  interactions  would  not  conform  to  the
 standard ULIRG$\rightarrow  $QSO emergence scenario,  though they can
 still ignite a starburst in  the gas-rich progenitor (Di Mateo et al.
 2007).  They may also be the observational answer to the expectations
 of  current  galaxy-formation theories  that  predict  the so  called
 ``wet-dry'' mergers  between gas-rich and gas-poor  progenitors to be
 the most common in the Universe (e.g.  Springel et al.  2005).

The  demanding combination of  observational tools  it took  to reveal
these intriguing  aspects of HE\,0450-2958  may come to  exemplify why
such interactions have been missed until now.  Moreover, it could also
resolve the conflicting views regarding  the host galaxies of the more
powerful optical  QSOs ($\rm M_v<-24$) in the  local Universe.  Unlike
the low luminosity AGNs such as Seyferts where a consistent picture of
AGNs  in  centers  of  gas-rich vigorously  star-forming  spirals  has
emerged  (e.g.   Schinnerer  et  al.   1998; Staghun  et  al.   2004),
detailed {\it HST}  studies by Dunlop et al.  2003  (see also Floyd et
al.   2004) conclude  that the  more powerful  AGNs reside  in typical
massive  ellipticals  ($\rm   R^{1/4}$  light  profiles,  old  stellar
populations).   On the  other hand  Scoville  et al.   2003 (see  also
Bertram et  al.  2007)  found large amounts  of molecular gas  in such
systems and concluded otherwise.

It is now clear that  the latter conclusions are premature. Indeed all
current  studies  of   the  hosts  of  luminous  AGN   lack  the  {\it
combination}  of:  a)  high  resolution  optical  imaging  along  with
specialized  deconvolution techniques  (allowing sensitive  probing of
the  host   galaxy  properties  against   the  bright  QSO),   and  b)
interferometric  CO  maps with  good  resolution,  high  S/N, and  the
excellent astrometry  that helped identify  the type of merger  in the
HE\,0450-2958 system.  Indicative of those limitations is PDS~456, the
most luminous QSO in the local Universe (also a ULIRG/QSO composite in
terms of its SED), where despite sensitive high resolution optical and
CO imaging observations, the QSO  host galaxy type and the location of
the  molecular gas  with respect  to its  AGN remain  unclear  (Yun et
al. 2004).

The short gas consumption timescales by a starburst, which are shorter
still for optically powerful  AGNs residing in gas-poor hosts, suggest
that these  twin ``beacons'' of  gas-rich/gas-poor galaxy interactions
may always be  in close proximity.  The advent  of sensitive mm/sub-mm
arrays,  along  with the  already  available  high resolution  optical
imaging capabilities now allows their study, and the ULIRG$\rightarrow
$QSO transition objects are excellent~candidates.

\section{Conclusions}

We performed  CO line emission  imaging observations of  the enigmatic
QSO/galaxy pair HE\,0450-2958  using the ATCA (CO J=1--0)  and the SMA
(CO J=3--2), detecting  the J=1--0 line and placing  an upper limit on
the J=3--2 line, our conclusions are as follows

\noindent
1. The CO J=1--0 line emission is associated with the companion galaxy
 not  the QSO, it  corresponds to  $\rm M_{gal}(H_2)  \sim (1-2)\times
 10^{10}\,  M_{\odot}$ and  amounts to  $\ga  30 \%$  of the  enclosed
 dynamical mass  within $\sim 8$\,kpc. This large  gas reservoir fuels
 extreme  star formation at  a rate  of $\rm  \sim 370\,  M_{\odot }\,
 yr^{-1}$,  efficiencies  of  $\rm L_{IR}/M_{gal}(H_2)\sim  (90-165)\,
 M_{\odot }/L_{\odot }$, and places  the companion galaxy on the upper
 gas-rich and star  forming end of ULIRGs.  This  property, along with
 the  disturbed   nature  of  the  system,   solidifies  the  strongly
 interacting status of this controversial galaxy/QSO pair.

\noindent
2. No  molecular gas  is detected  towards  the QSO  itself with  $\rm
  M_{QSO-host}(H_2)/M_{gal}(H_2)\leq 1/5$,  consistent with the little
  dust extinction or star  formation activity deduced for its putative
  host galaxy.   This weakens the case  for a host  galaxy typical for
  Narrow Line  Seyfert 1  AGNs since these  are gas-rich  spirals with
  luminous circumnuclear starbursts.   HE\,0450-2958 is thus the first
  known  case of  a gas-rich  extreme  starburst and  a powerful  QSO,
  intimately linked (via a strong interaction), but not~coincident.

\noindent
3. Most of the radio continuum in this system is due to low-brightness
   AGN-driven jets rather than  star formation activity.  Deeper radio
   continuum observations  are needed  to reveal their  morphology and
   extent, which in turn can yield more clues regarding its AGN.

\noindent
4. The early  prominence of HE\,0450-2958  as an IR-selected  QSO that
  exemplifies   the   ULIRG$\rightarrow   $(optically  luminous   QSO)
  transition  in  a  popular  evolutionary scenario,  and  its  hereby
  revealed nature  as a  strong interaction between  a gas-rich  and a
  gas-poor, possibly elliptical, galaxy  opens up the possibility that
  some  of these  so-called  transition objects  are such  interacting
  pairs  instead.  Such   gas-poor/gas-rich  galaxy  interactions  can
  activate a  starburst in the  gas-rich progenitor, and  an optically
  luminous AGN  in the gas-poor  one.  Their composite  dust continuum
  would then  be typical of IR-warm  ULIRG$\rightarrow $QSO transition
  objects.  Sensitive  optical and CO  imaging at high  resolution can
  now  reveal  and  study   such  dynamical  interactions,  and  these
  transition objects are excellent candidates.

\section{Acknowledgments}

Jeff Wagg is grateful for  support from the Max-Planck Society and the
Alexander von Humboldt Foundation. The anonymous referee is gratefully
acknowledged  for  numerous  suggestions  that greatly  clarified  the
original  document.  We  also thank  all the  people that  operate and
maintain  ATCA  and  in  particular  its  3\,mm  receivers  for  their
dedicated  support.   Padelis  Papadopoulos  thanks  Karl  Jesienowski
aboard  the  {\it  Undersea  Explorer} for  great  conversations,  and
Marcella  Carollo  for  bringing  this  controversial  object  to  his
attention.  Last  but not  least  he  would  like to  thank  Margarita
Zakalkas  for inspiration during  his time  in Zurich.   The Australia
Telescope is funded by the  Commonwealth of Australia for operation as
a National Facility managed by CSIRO.

\clearpage

\begin{figure}[h]
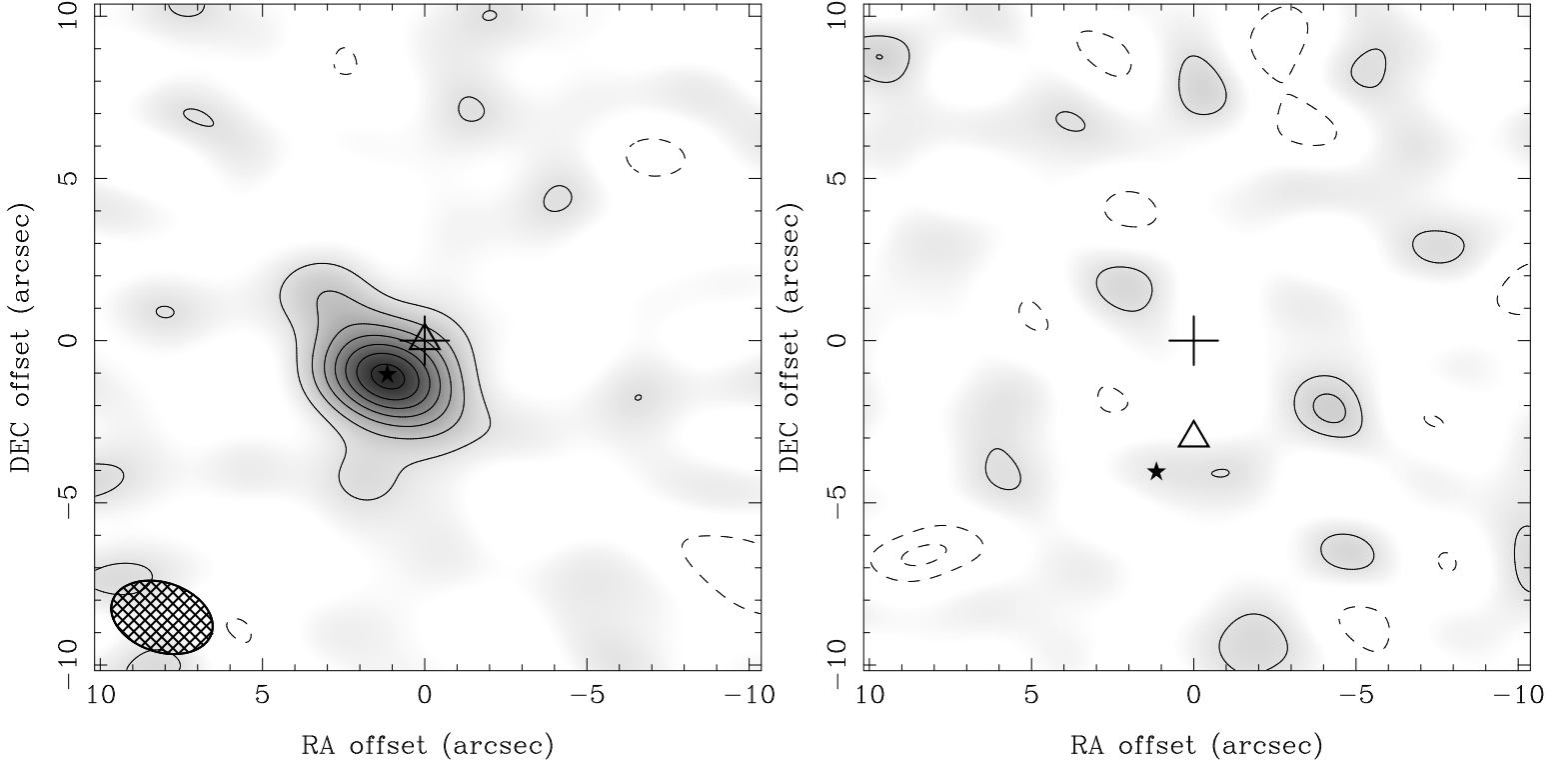

\vspace*{12cm}
\includegraphics{./f1a.ps}
\includegraphics{./f1b.ps}
\caption{{\it Left:} Map of  CO J=1--0 emission (NA-weighted, CLEANed)
 averaged over $\rm 570\,km\,s^{-1}$  (see section 2.3), and contours:
 $\rm (-2,2,4,6,8,10,12,14)\times  \sigma_{rms}$. {\it Right:}  The mm
 continuum emission at  94.53\, GHz with contours of  $\rm (-3, -2, 2,
 3)\times  \sigma_{rms}$.   The noise  in  both  maps  is $\rm  \sigma
 _{rms}\sim 0.45\, mJy/beam$,  and the restoring beam is  shown at the
 bottom  left  of  the  CO  image  ($\rm  \Theta  _{beam}=3.21''\times
 2.14''$,  $\rm PA=72^{\circ}$).  The  phase center  is marked  by the
 cross, the  diamond marks  the AGN position,  and the star  marks the
 peak of the CO emission (see  Table 1).  For the mm continuum imaging
 the phase center was set $3''$  to the north of the AGN's position to
 avoid  any  DC correlator  offsets  ``masking''  as  a spurious  weak
 source.}
\end{figure}

\clearpage

\begin{figure}
\includegraphics[angle=0,scale=0.8]{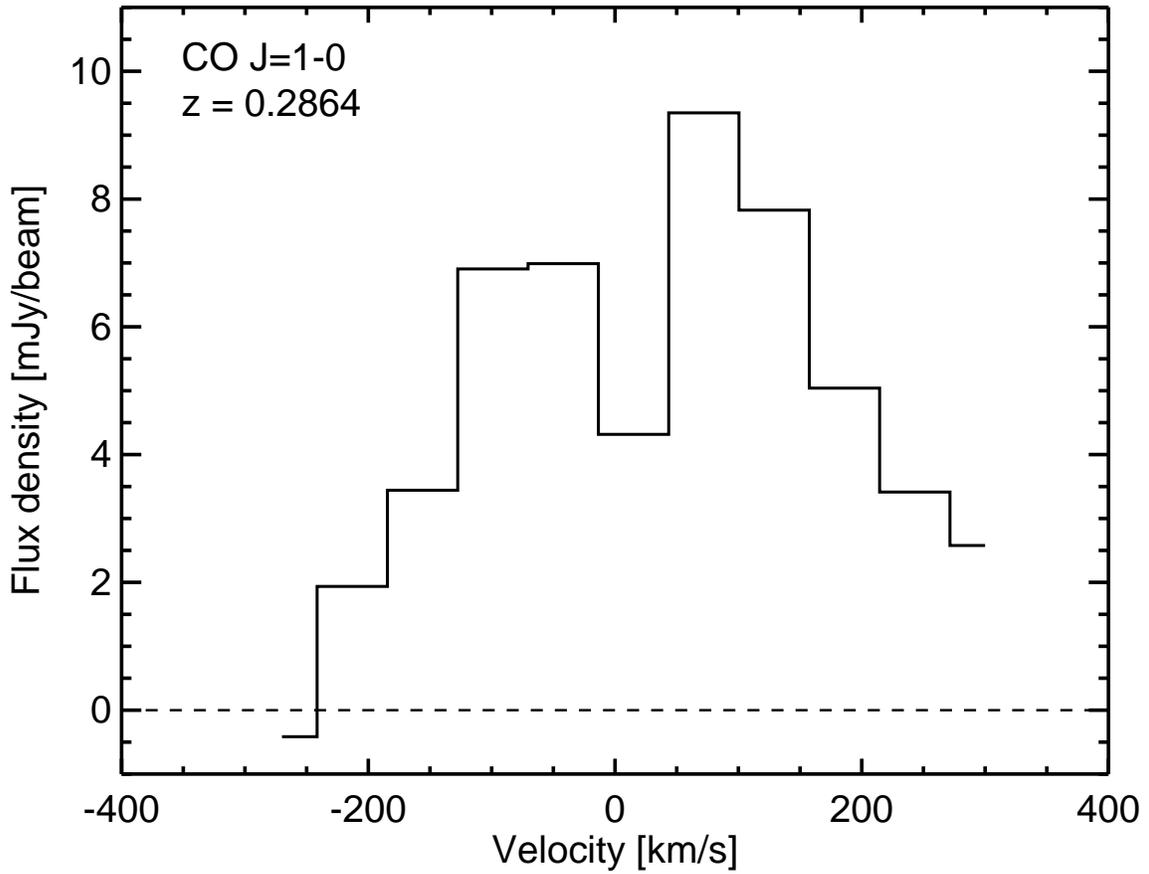}
\caption{The  spectrum  corresponding  to  the  peak  CO  J=1--0  line
emission (Figure  1). The velocity range is centered on redshift of the
 QSO/ULIRG system shown on the upper left.}
\end{figure}

\clearpage

\begin{figure}
\includegraphics[angle=0,scale=0.8]{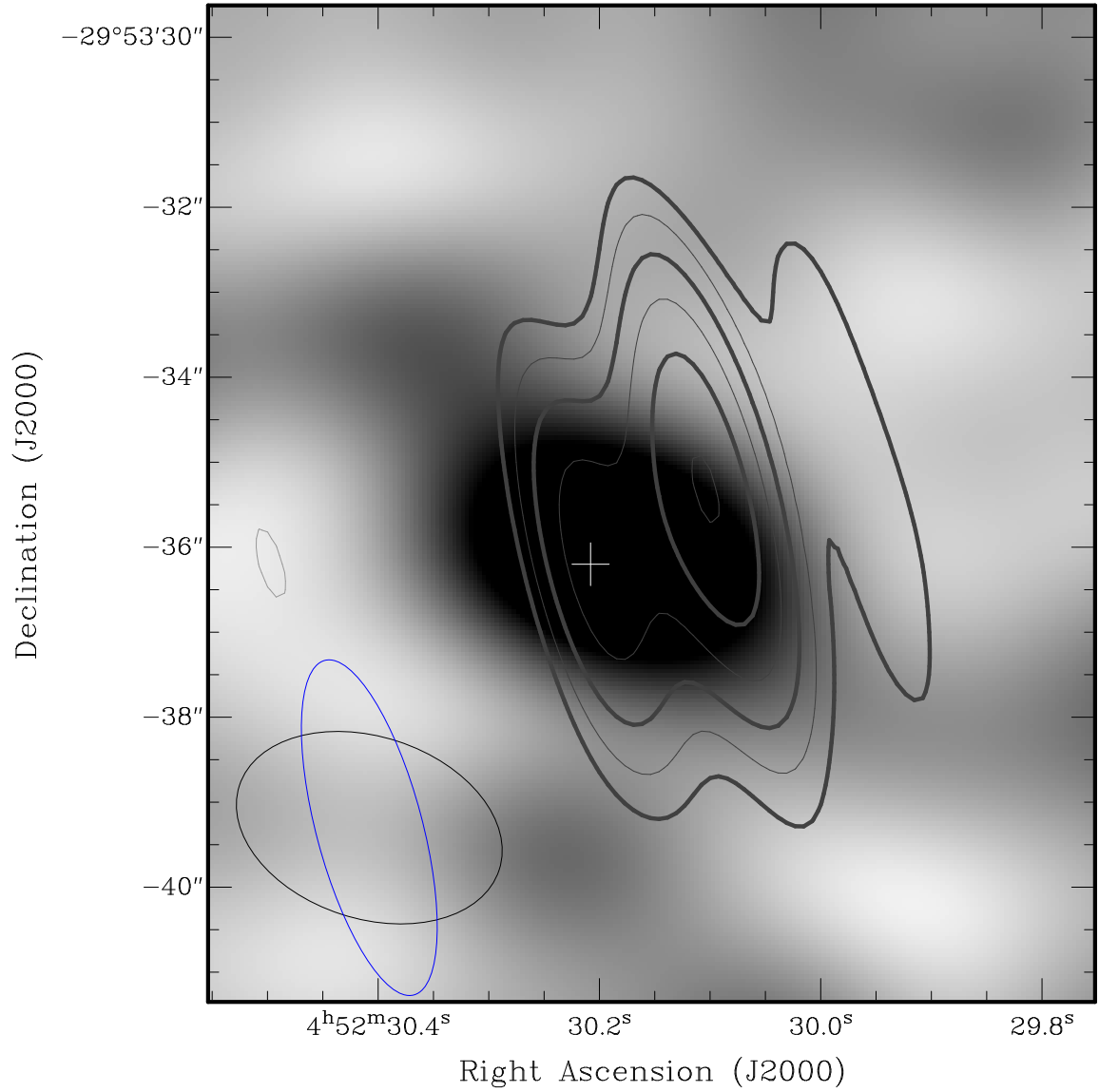}
\caption{ATCA CO(1-0) greyscale image with ATCA 8.4GHz radio continuum
contours overlaid. Contour  levels start at $200\mu$Jy~beam$^{-1}$ and
increase in a geometric series  with a common ratio of $\sqrt{2}$. The
CO(1-0) and the 8.4\,GHz beams are  shown in the bottom left corner of
the  image.  The cross  marks  the peak  emission  of  the C\,2  radio
continuum emitting region associated  with the companion galaxy (Feain
et al. 2007).}
\end{figure}

\clearpage

\begin{figure}[h]
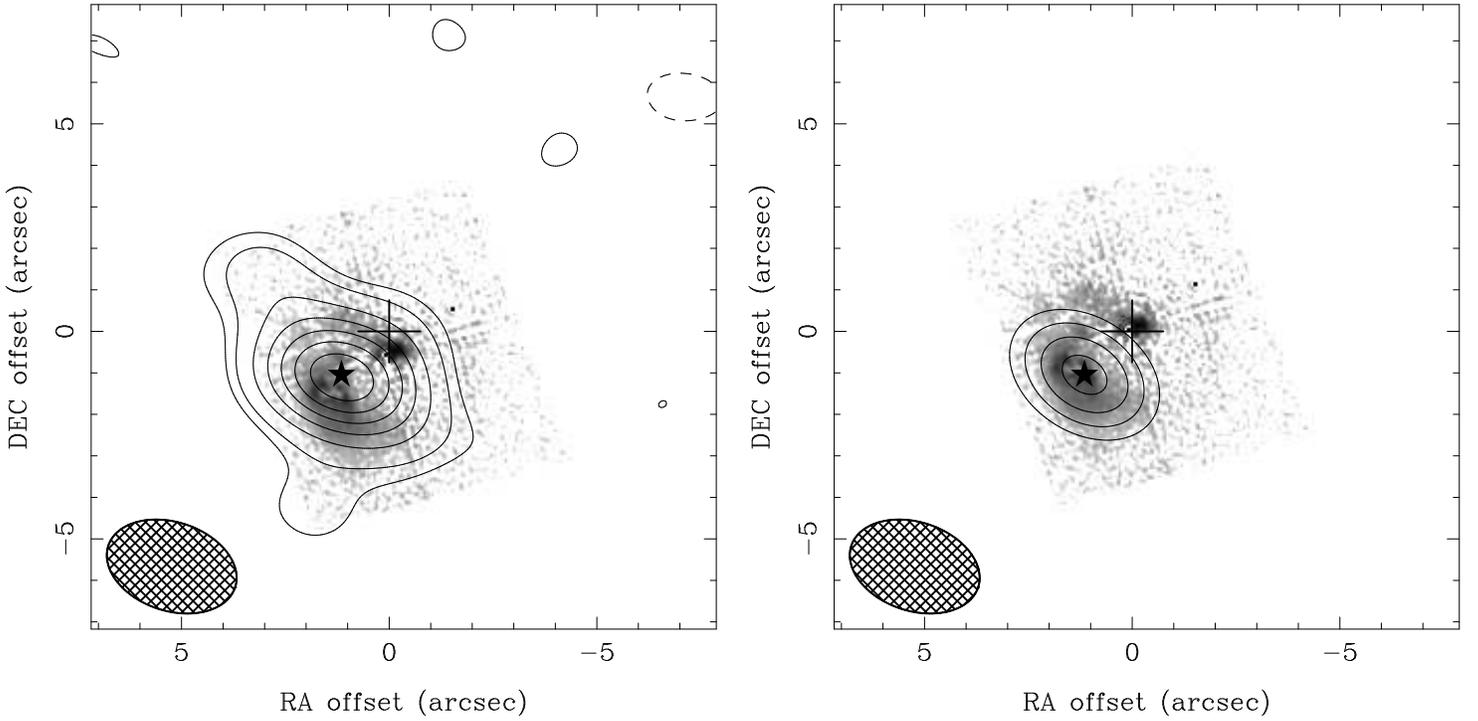

\vspace*{12cm}
\includegraphics{./f4a.ps}
\includegraphics{./f4b.ps}
\caption{{\it Left:}  CO J=1--0  emission overlaid on  the deconvolved
  HST/ACS image of the HE\,0450-2958 system, with the QSO contribution
  subtracted (from Magain  et al.  2005).  Contours are:  $\rm (-2, 2,
  5,7,9,11,13)\times       \sigma       _{rms}$,       with       $\rm
  \sigma_{rms}=0.45\,mJy/beam$. The  beam is shown at  the bottom left
  (HPBW=$3.21''\times 2.14''$, $\rm  PA=72^{\circ}$). {\it Right:} The
  {\it deconvolved}  CO source model  (see section 2.3.1)  overlaid on
  the HST  image after a $\Delta \delta_{o}=0.6''$  shift aligning the
  optical and radio AGN positions.   The good correspondence of the CO
  emission with the companion  ULIRG is~evident.  In the optical image
  the QSO (marked by the single bright pixel near the phase center) is
  surrounded by an AGN-excited gas cloud.  The cross size at the phase
  center   $(0'',0'')$   denotes   the  (dominant)   HST   astrometric
  uncertainties ($\rm \sim 0.5''-1.0''$), and the star symbol denotes
  the peak CO J=1--0 emission.}
\end{figure}

\clearpage

\begin{figure}
\includegraphics[angle=0,scale=0.9]{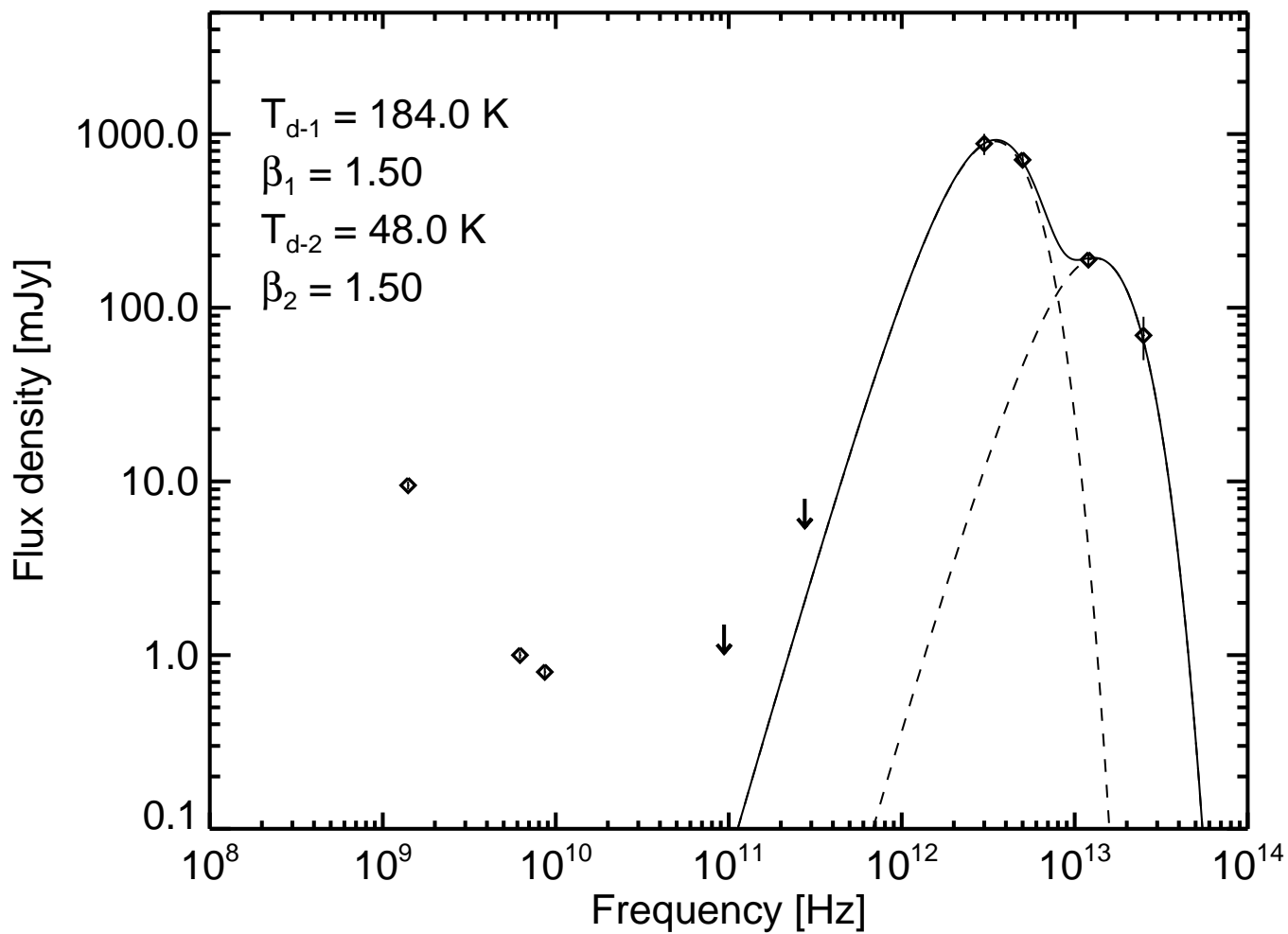}
\caption{Two   component  fit   of   the  mm/IR   dust  continuum   of
HE\,0450-2958 (see  section 3.2). The  upper limits at  mm wavelengths
were obtained  from this work  (see Table 1),  and the IRAS  fluxes at
$12\,\mu$m, $25\,\mu$m, $60\,\mu$m, and $100\,\mu$m are from Moshir et
al. 1990.  The emissivity laws adopted and  the resulting temperatures
from the  2-component dust  emission fit are  shown in the  upper left
(see also 3.2).}
\end{figure}

\clearpage

\begin{deluxetable}{lcc}
\tablecolumns{3}
\tablewidth{0pc}
\tablecaption{Properties of the HE\,0450--2958 system}
\tablehead{
\colhead{Property} & \colhead{AGN} & \colhead{Companion galaxy} }
\startdata
$\rm \alpha (J2000.0)$ & 04$\rm ^h$ 52$\rm ^m$ 30$\rm ^s$.1 & $+1.15''\pm 0.12''$\,\tablenotemark{a}\\
$\rm \delta (J2000.0)$ & --29$^{\circ}$ 53$'$ 35.0$''$ & $-1.05''\pm 0.10''$\,\tablenotemark{a}\\
$z$\,\tablenotemark{b}    & 0.2863 & 0.2865 \\
$\rm S_{peak}$(1--0) (mJy/beam)\tablenotemark{c}  &   $ <1.35  $  & $6.7\pm 0.45$\\
$\rm S_{peak}$(3--2) (mJy/beam)\tablenotemark{c}  &   $ < 20   $  & $ < 20 $\\
$\rm \int  S_{CO(1-0)} dV$ (Jy\,km\,s$^{-1}$)\tablenotemark{c} & $<0.77 $ & $5.70\pm1.15$\\
$\rm \int  S_{CO(3-2)} dV$ (Jy\,km\,s$^{-1}$)\tablenotemark{c} & $<11$ & $<20 $\\
Intrinsic CO(1-0) source size & \nodata & $2.5''\times 1.5''$\\
$\rm M(H_2)$ $\rm (\times 10^{10}\,M_{\odot })$\,\tablenotemark{d} &  \nodata  &  $1.25-2.3$\\
$\rm S_{94\,GHz}$ (mJy)\tablenotemark{c}  & $< 1.4$ & $ < 1.4$\\
$\rm S_{275\,GHz}$ (mJy)\tablenotemark{c}  & $< 8$ & $ < 8$\\
$\rm L_{IR}$  $\rm (\times 10^{12}\,L_{\odot})$\tablenotemark{e}    & 2.6  & 2.1 \\
\enddata
\tablenotetext{a}{Position of the peak CO J=1--0 emission with respect to the\\
\hspace*{0.5cm} AGN (see 2.3.1)}
\tablenotetext{b}{Canalizo \& Stockton 2001}
\tablenotetext{c}{All limits are $3\sigma$, point source assumed for AGN-related limits}
\tablenotetext{d}{See section 3}
\tablenotetext{e}{from the dust continuum SED fit in Figure 6}

\end{deluxetable}

\end{document}